\documentclass{article}
\begin{document}
\title{Linearisation of optical effects at low light levels.
\author{Jacques. Moret-Bailly 
\footnote{Laboratoire de physique, Université de Bourgogne, BP 47870, F-21078 
Dijon cedex, France.
email : jmb@jupiter.u-bourgogne.fr
}}}
\maketitle

\begin{abstract}
As a light beam is produced by an amplification of modes of the zero point field in its source, this 
field cannot be distinguished; consequently a nonlinear optical effect is a function of the total field. 
However, we generally prefer to use a conventional field which excludes the zero point field; for a 
low conventional field, the total field may be developed to the first order, so that the effect appears 
linear.

This nearly trivial remark allows a correct computation of the signal of a photocell used for 
photon counting and shows that the "impulsive stimulated Raman scattering" (ISRS), a nonlinear, 
without threshold effect, which shifts the frequencies, becomes linear at low light levels, so that the 
shifted spectra are not distorted.

\medskip
Comme un faisceau de lumière résulte d'une amplification de modes du champ du point zéro par 
sa source, le champ du point zéro ne peut être distingué ; en conséquence, un effet optique non-
linéaire est une fonction du champ total. En prenant la définition usuelle du champ qui exclut le 
champ du point zéro, pour un champ usuel faible le champ total peut être développé au premier 
ordre, de sorte que l'effet devient linéaire.

Cette remarque quasiment triviale permet, en particulier, de rendre compte correctement de la 
détection de la lumière d'un photorécepteur en ''comptage de photons'', et de montrer que la 
diffusion Raman impulsionnelle stimulée, effet quadratique sans seuil, se transforme, à bas niveau 
en un effet linéaire qui fait glisser les fréquences spectrales sans distordre les spectres.

pacs{42.25Bs, 42.50Gy}

\end{abstract}

\maketitle
%

\section{Introduction}
Introduced by quantum electrodynamics, the zero point electromagnetic field appears as a strange physical 
concept. The transformation of the first (wrong) Planck's law into the second \cite{Planck,Nernst} sets its 
value $h\nu/2$, but not its nature. Stochastic electrodynamics \cite{Marshall} describes the zero point 
field, renamed "stochastic field" as an ordinary field, but the strongest (although qualitative) interpretation, is 
provided by the old classical theory : The electric field radiated by an oscillating electric dipole is known; if 
there is no external field, the dipole is a source; but if it is merged in an external field of the same frequency, 
with convenient polarisations and phases, it partly cancels the external electromagnetic field, decreasing the 
electromagnetic energy : the dipole is a receiver ; as a large part of the fields is not cancelled, the dipole not 
only absorbs a part of the incident field, it scatters it. Thus, the absorption of the field emitted by a dipole 
requiring an infinite number of dipoles, it exists a stochastic unabsorbed, scattered field. This description 
shows that the zero point field is an ordinary field.
The measure of the Einstein coefficients $A$ and $B$ for the spontaneous and stimulated emissions shows 
that the spontaneous emission is exactly induced by the zero point field. Thus the field in a light beam is a 
zero point field amplified by a source, and it is artificial to distinguish in it a zero point field and the 
remainder, the field radiated spontaneously in the old theory (thereafter the conventional field). Thus the 
conventional field has no physical existence, it must not appear in the formula describing an optical effect.
\section{Absorption and detection}
Usually, we write that the intensity absorbed or detected by a photocell is proportional to the square of the 
amplitude of the conventional electric field, this square being considered proportional to the flux of 
electromagnetic energy. This supposes that there is no coherence between the conventional field and the 
stochastic field, an assumption which is false. How can we write that in the dark there is no absorption 
while the stochastic intensity hits a photoelectric cell ? A solution is supposing that there is an equilibrium 
between the absorbed stochastic field and a reemission. Remark that in cold, good photocells it remains a 
signal which seems produced by the long and powerful enough fluctuations of the stochastic field. $E_0$ 
being the amplitude in a mode of the stochastic field and $\beta E_0$ the field resulting of an amplification 
of this mode by a source, the net available energy on a receiver is $(\beta E_0)^2-E_0^2=2(\beta-
1)E_0^2+((\beta-1)E_0)^2$.

If $\beta$ is nearly one, the second term may be neglected ; for a given optical configuration, the time-
average of the stochastic amplitude $E_0^2=|E_0|^2$ is constant, so that {\it the detected signal is 
proportional to the amplitude of the conventional field}. On the contrary, for a high conventional field, the 
usual rule is got.

This result is experimentally verified by the fourth order interference experiments with photon counting 
(see, for instance, \cite{Clauser,Gosh,Ou1,Ou2,Kiess}). The result of these experiments is easily got {\it 
qualitatively} using the classical rules \cite{M942}, but the contrast of the computed fringes is lower than 
shown by the experiments. In the simplest experiment \cite{Gosh} two small photoelectric cells  are put in 
the interference fringes produced by two point sources; the interferences are not visible because they 
depend on the fast changing difference of phase $\phi$ of the sources. The sources are weak; the signal is 
the correlation of the counts of the cells.

Distinguishing the photoelectric cells by an index j equal to 1 or 2, set $\delta_j$ the difference of paths for 
the light received by the cells. The amplitude of the conventional field received by a cell is proportional to 
$\cos (\pi\delta_j/\lambda+\phi/2)$, so that, assuming the linearity, the probability of a simultaneous 
detection is proportional to
\begin{equation}
 \cos (\frac{\pi\delta_1}{\lambda}+\frac{\phi}{2})\cos (\frac{\pi\delta_2}{\lambda}+\frac{\phi}{2}).
\end{equation}
 The mean value of this probability got by an integration over $\phi$ is zero for $\delta_1-
\delta_2=\lambda/2$, so that the visibility has the right value 1. Assuming the usual response of the cells 
proportional to the square of the conventional field, the visibility would have the wrong value 1/2.

\section{Low level "Impulsive Stimulated Raman Scattering" (ISRS).}
ISRS, known since 1968 \cite{Yan} is now commonly used \cite{Nelson}. It is not a simple Raman 
scattering, but a parametric effect, combination of two {\it space-coherent} Raman scattering, so that the 
state of the interacting molecules is not changed. ISRS is obtained using ultrashort light pulses, that is 
"pulses shorter than all relevant time constants" \cite{Lamb}, usually femtosecond laser pulses. In a gas, 
the relevant time constants are:

i) the collisional time : the collisions destroy the coherence of the excitation of the molecules.

ii) the period which corresponds to the virtual Raman transition : the scattered light interferes with the 
exciting light into a frequency-shifted single beam so that the time-coherence of the output beams is 
not broken by the dispersion and the effect is strong.

ISRS is generally performed using at least a strong pump laser beam so that it is nonlinear, the frequency 
shift depends on the intensity of the beam. But it has no threshold : a direct study \cite{Moret,Moret2} 
shows what happens if the pump beams are usual incoherent light beams, made of relatively long, weak 
pulses : the effect becomes linear so that the relative frequency shift $\Delta\nu/\nu$ depends slightly on a 
dispersion, not the intensity. The coherence preserves the wave-fronts; thus there is no blur either in the 
images or in the spectra, just as by a Doppler frequency shift. Thought the coherence of the effect called 
"Incoherent Light Coherent Raman Scattering" (ILCRS) makes it strong, it requires so low pressures that it 
seems impossible to perform it in the labs. The Universe, however provides good experimental conditions : 
the paths may be long, a lot of mono- or poly-atomic molecules have hyperfine structures providing the 
low energy Raman transitions : atoms perturbated by Zeeman effect near the quasars, H$_2^+$ molecules 
in the clouds detected by the forbidden nuclear spin transition of H$_2$ at 0.2m\dots

A part of the redshifts attributed to Doppler (or expansion) effect is surely provided by ILCRS able to 
transfer energy from high frequencies to isotropic thermal radiation (2.7K). Near bright stars, this transfer 
may be similar to a transfer by heated dust.

\section{Conclusion}
The nonlinear light-matter interactions without threshold become linear using weak light beams. In two 
examples, this trivial property provides an interesting expansion of well known effects; it explains many 
other effects, for instance the computation of the sub-Poissonian statistics in photon counting \cite{Short}
is easier than the quantum computation \cite{Glauber} in particular in an intermediate case where the light 
flux is too large.

\end{document}